\documentclass[preprint2]{aastex6}
\usepackage{xcolor}
\usepackage[utf8]{inputenc}

\newcommand\vsini{$v$\,sin\,$i_\star$}

\newcommand\vmic{$V_{\mathrm{mic}}$}
\newcommand\vmac{$V_{\mathrm{mac}}$}

\newcommand\teff{$T_{\rm{eff}}$}

\newcommand\logg{{\it log}\,{\it g$_\star$}}
\newcommand\loggp{{\it log}\,{\it g$_p$}}

\newcommand\Msun{\hbox{$M_{\odot}$}}  %Msun
\newcommand\Rsun{\hbox{$R_{\odot}$}}  %Rsun
\newcommand\Mjup{\hbox{$M_\mathrm{Jup}$}}  %Mjup
\newcommand\Rjup{\hbox{$R_\mathrm{Jup}$}}  %Rjup
\newcommand\kms{\hbox{km\,s$^{-1}$}}  %km per sec
\newcommand\ms{\hbox{m\,s$^{-1}$}}  %km per sec
\newcommand\gcm{\hbox{g\,cm$^{-3}$}}  % g/cm^3

\newcommand\snameone{K2-60}
\newcommand\pnameone{K2-60b}

\newcommand\snametwo{C7\_8514}
\newcommand\pnametwo{C7\_8514b}
\newcommand\pnametwoa{EPIC 216468514b}

\newcommand\sradone{$1.12 \pm 0.05 $}
\newcommand\smassone{$0.97 \pm 0.07 $}
\newcommand\steffone{$5500 \pm 100$}
\newcommand\sloggone{$4.07 \pm 0.11$}
\newcommand\slogglcone{$4.33 \pm 0.04$}
\newcommand\smetone{$0.01 \pm 0.11$}
\newcommand\pradone{$0.683 \pm 0.037$}
\newcommand\pmassone{$0.426 \pm 0.037$}
\newcommand\pteffone{$1400 \pm 50 $}
\newcommand\ploggone{$3.35 \pm 0.06$}
\newcommand\sageone{$ 10.0 \pm 3.0 $}

\newcommand\sradtwo{$1.78 \pm 0.16 $}
\newcommand\smasstwo{$1.30 \pm 0.14 $}
\newcommand\stefftwo{$6030 \pm 120$}
\newcommand\sloggtwo{$4.07 \pm 0.10$}
\newcommand\slogglctwo{$4.05 \pm 0.07$}
\newcommand\smettwo{$0.10 \pm 0.10$}
\newcommand\pradtwo{$1.44 \pm 0.15$}
\newcommand\pmasstwo{$0.84 \pm 0.08$}
\newcommand\ptefftwo{$1780 \pm 90 $}
\newcommand\ploggtwo{$3.00 \pm 0.07$}
\newcommand\sagetwo{$4.25 \pm 1.75$}

\definecolor{red}{rgb}{1,0,0}

\newcommand\footnoteref[1]{\protected@xdef\@thefnmark{\ref{#1}}\@footnotemark}

\shorttitle{\pnameone\ and \pnametwoa. A Sub-Jovian and a Jovian Planet from the K2 mission}
\shortauthors{Eigm\"uller et al.}

\begin{document}

\title{\pnameone\ and \pnametwoa. A Sub-Jovian and a Jovian Planet from the K2 mission}

\author{Philipp Eigm\"uller\altaffilmark{1},
Davide Gandolfi\altaffilmark{2,3}, 
Carina M. Persson\altaffilmark{4},
Paolo Donati\altaffilmark{5},
Malcolm Fridlund\altaffilmark{4,6},
Szilard Csizmadia\altaffilmark{1}, 
Oscar Barrag\'an\altaffilmark{2},
Alexis M. S. Smith\altaffilmark{1},
Juan~Cabrera\altaffilmark{1}, 
Judith Korth\altaffilmark{7}, 
Sascha Grziwa\altaffilmark{7}, 
Jorge~Prieto-Arranz\altaffilmark{8,9},
David~Nespral\altaffilmark{8,9}, 
Joonas Saario\altaffilmark{10}
William~D.~Cochran\altaffilmark{11},
Felice~Cusano\altaffilmark{4},
Hans J. Deeg\altaffilmark{8,9}, 
Michael~Endl\altaffilmark{11},
Anders Erikson\altaffilmark{1}, 
Eike~W.~Guenther\altaffilmark{12}, 
Artie~P.~Hatzes\altaffilmark{12}, 
Martin~P\"atzold\altaffilmark{7}, 
and Heike~Rauer\altaffilmark{1,13}}

\altaffiltext{1}{Institute of Planetary Research, German Aerospace Center, Rutherfordstrasse 2, 12489 Berlin, Germany}
\altaffiltext{2}{Dipartimento di Fisica, Universit\'a di Torino, via P. Giuria 1, 10125 Torino, Italy}
\altaffiltext{3}{Landessternwarte K\"onigstuhl, Zentrum f\"ur Astronomie der Universit\"at Heidelberg, K\"onigstuhl 12, 69117 Heidelberg, Germany}
\altaffiltext{4}{Department of Earth and Space Sciences, Chalmers University of Technology, Onsala Space Observatory, 439 92 Onsala, Sweden}
\altaffiltext{5}{INAF – Osservatorio Astronomico di Bologna, Via Ranzani, 1, 40127, Bologna}
\altaffiltext{6}{Leiden Observatory, University of Leiden, PO Box 9513, 2300 RA, Leiden, The Netherlands}
\altaffiltext{7}{Rheinisches Institut f\"ur Umweltforschung an der Universit\"at zu K\"oln, Aachener Strasse 209, 50931 K\"oln, Germany}
\altaffiltext{8}{Instituto de Astrof\'isica de Canarias, 38205 La Laguna, Tenerife, Spain}
\altaffiltext{9}{Departamento de Astrof\'isica, Universidad de La Laguna, 38206 La Laguna, Spain}
\altaffiltext{10}{Nordic Optical Telescope, Apartado 474, 38700, Santa Cruz de La Palma, Spain}
\altaffiltext{11}{Department of Astronomy and McDonald Observatory, University of Texas at Austin, 2515 Speedway, Stop C1400, Austin, TX 78712, USA}
\altaffiltext{12}{Th\"uringer Landessternwarte Tautenburg, Sternwarte 5, D-07778 Tautenberg, Germany}
\altaffiltext{13}{Center for Astronomy and Astrophysics, TU Berlin, Hardenbergstr. 36, 10623 Berlin, Germany}

% TODO
% stellar classification
% limit on eccentricity of first planet
% discussion of second planet
% do one final characterization run
% read everything
% is \snametwo a main sequence star
% reported epoch of star2 needs correction

\begin{abstract}
We report the characterization and independant detection of \pnameone, as well as the detection and characterization of EPIC~216468514b, two transiting hot gaseous planets from the K2 space mission. We confirm the planetary nature of the two systems and determine their fundamental parameters combining the K2 time-series data with FIES@NOT and HARPS-N@TNG spectroscopic observations. \pnameone\ has a radius of \pradone \Rjup\ and a mass of \pmassone \Mjup\ and orbits a G4\,V star with an orbital period of $3.00267\pm0.00006$~days. EPIC~216468514b has a radius of \pradtwo \Rjup\  and a mass of \pmasstwo \Mjup\ and orbits an F9\,IV star every $3.31392 \pm 0.00002$ days. \pnameone\ is among the few planets at the edge of the so-called ``desert'' of short-period sub Jovian planets. EPIC~216468514b is a highly inflated Jovian planet orbiting an evolved star about to leave the main sequence.
%Both systems have been photometrically monitored by the K2 space mission during its Campaign 3 and Campaign 7, respectively. 
%Transits of \pnameone and \pnametwo were observed during the extended mission of the Kepler spacecraft, K2 in Campaign 3 and Campaign 7, respectively. One of the planets, \pnameone, has already been published as validated planet wit unknown mass. 
\end{abstract}

\keywords{Planetary systems – planets and satellites: detection – planets and satellites: gaseous planets – planets and satellites: individual: \snameone\,b, EPIC~216468514\,b.}

\section{Introduction}

More than $3,500$ exoplanets have been discovered over the last 25 years \citep{Schneider2011}\footnote{From \url{http://www.exoplanet.eu}, as of 26 September 2016.}. This has allowed us to compare the observed exoplanet populations with formation theories and evolutionary models \citep[e.g.,][]{Mordasini2009A, Mordasini2009B, Alibert2011, Mordasini2012}. 
One of the highly-discussed topics in exoplanetary science is the so called ``sub-Jovian desert'', which describes a significant dearth of exoplanets with masses lower than $\sim$300 Earth masses and orbital periods below 2-4 days
\citep{Szabo2011, Beauge2013, Mazeh2016}.

Whereas lower mass planets get reduced in size due to photo-evaporation \citep{Lundkvist2016}, hot Jovian planets, more massive than Jupiter and with orbital periods below 4 days, tend to be inflated. A detailed empirical study of these radius anomalies was conducted by \citet{Laughlin2011} who found a clear correlation between the planets' orbit-averaged effective temperatures and the observed inflation.  \citet{Laughlin2011} suggested that the Ohmic heating might account for the observed inflation. This effect could influence the upper border of this ``desert'' related to the radius. But as \citet{Mazeh2016} showed, this desert is also present in the mass regime. Recent theoretical studies on planet formation give additional explanations for the boundaries of the desert using \emph{in situ} formation \citep{Batygin2016}, as well as planet migration theories \citep{Matsakos2016}. Unfortunately, the lack of well characterized planets in the regime close to the sub-Jovian desert does not allow us at the moment to give strict constraints on its border. The upper border seems to be well defined due to the large amount of planets detected with ground based transit surveys, but, as a comparison with Kepler planets shows, the detection bias of these ground based surveys  does not allow to extrapolate the upper border of the sub Jovian desert to a regime for planets smaller than 0.8 \Rjup. The number of well characterized Kepler planets on the other hand are also very limited. A better empirical definition of the sub Jovian desert and its boundaries might allow further constraints to be placed on planet formation and evolution models.

Here we report our results on \pnameone\ and \pnametwo, both short period planets with orbital periods of $\sim$3 days. The small mass and size of \pnameone\ puts this planets close to the sub Jovian desert and thus might help to better restrict its boundaries in future. \pnametwo\ on the other hand is an highly inflated planet. It is a member of the inflated hot Jupiters, but is only one of few orbiting a sub Giant host star.

Planet \pnameone\ has been recently reported as a planet candidate by \citet{Crossfield2016} and validated using high resolution imaging by \citet{Schmitt2016}. However, the planet has not been characterized before in terms of mass and bulk density.

\section{Observations}
\subsection{K2 photometry and transit detection}

The Kepler space observatory, launched in 2009, was designed to provide precise photometric monitoring of over $150,000$ stars in a single field and to detect transiting Earth-sized planets with orbital periods up to one year \citep{2010Sci...327..977B}. In spring 2013, after 4 years of operation in space, the failure of the second reaction wheel caused the end of the mission, as it was not longer possible to precisely point the telescope. At the end of 2013 the operation of the Kepler space telescope re-started with a new concept that uses the remaining reaction wheels, the spacecraft thrusters, and Solar wind pressure, to point the telescope. The new mission, called K2 \citep{K2}, enables the continued use of the Kepler spacecraft with limited pointing accuracy. In contrast to the Kepler mission, K2 observes different fields located along the ecliptic for a duration of about three consecutive months per field. EPIC~206038483 (\snameone) was observed by the K2 mission in campaign 3 from 2014 November until 2015 February. EPIC~216468514 (\snametwo) was observed in campaign 7, between 2015 October and December.

\begin{figure}
\resizebox{0.5\textwidth}{!}{\plottwo{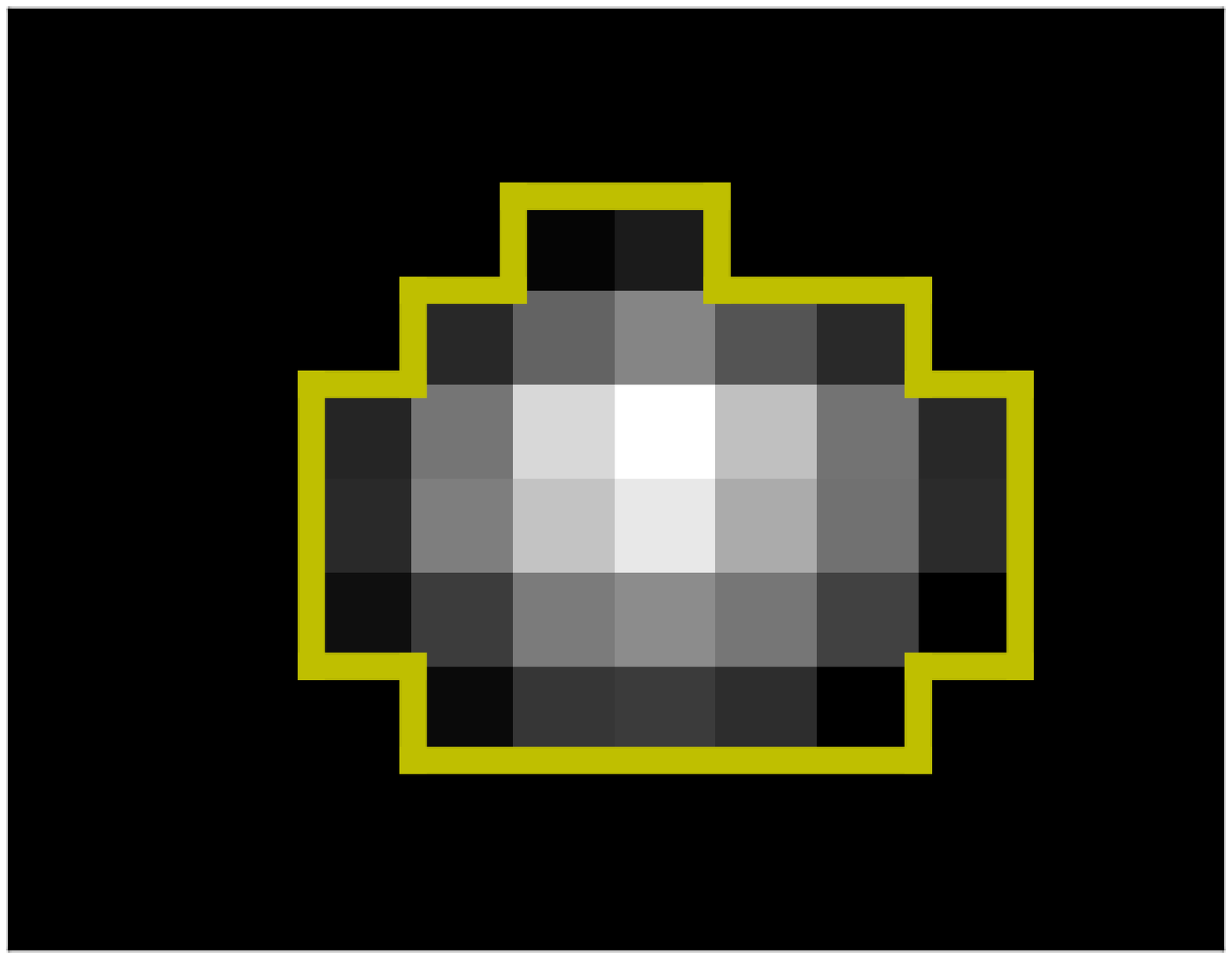}{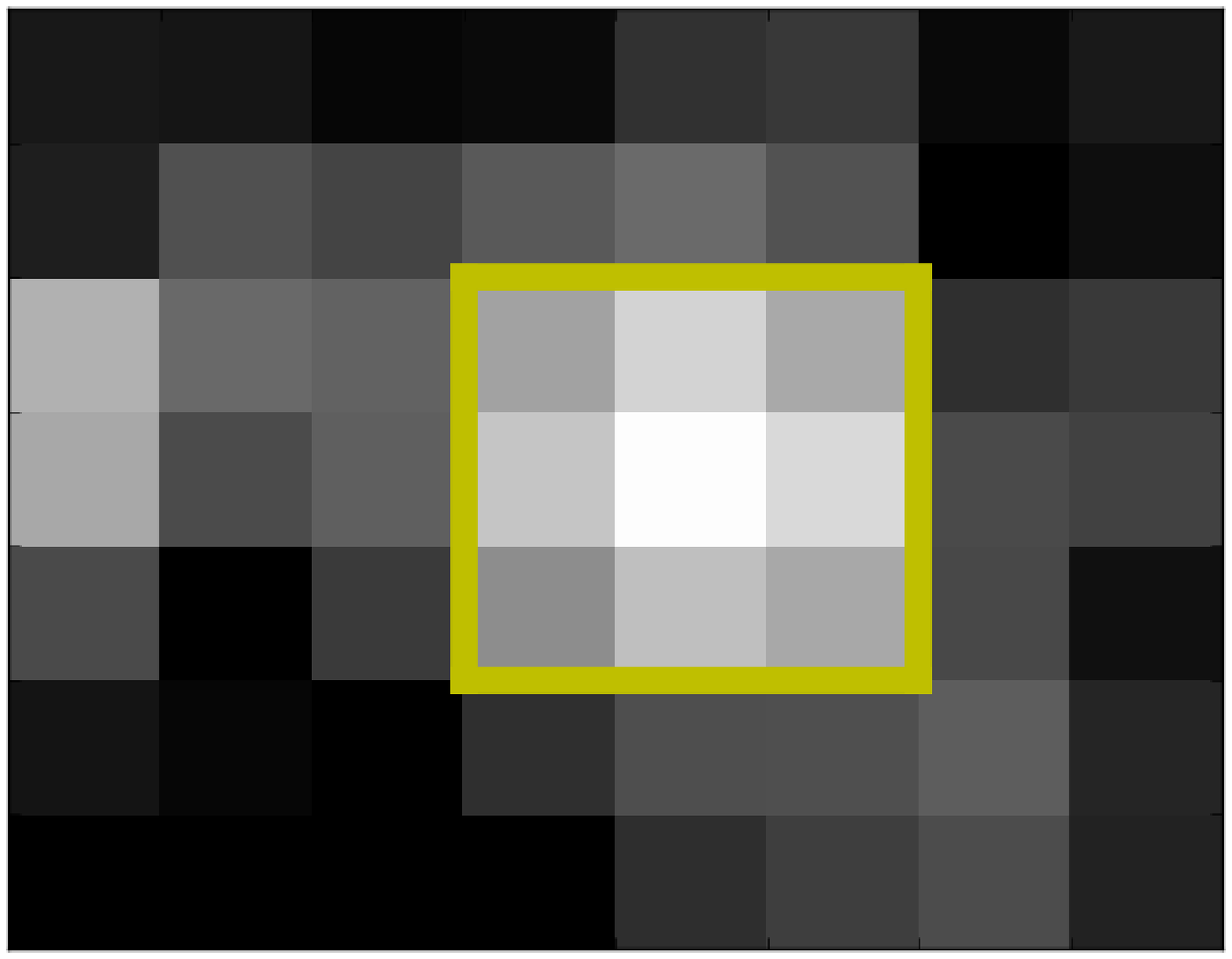}}
\caption{K2 stamps of \snameone\ (left) and   \snametwo\ (right).  The yellow lines represent the adopted photometric apertures. The pixel scale of the Kepler spacecraft is $3.98\arcsec$ per pixel. The stamp of \snameone\ has a size of 12x10 pixels, whereas the stamp of \snametwo\ has a size of 8x7 pixels. The gray scale represents the counts per pixel. \label{fig1}}
\end{figure}

\begin{figure}
\resizebox{0.5\textwidth}{!}{\plotone{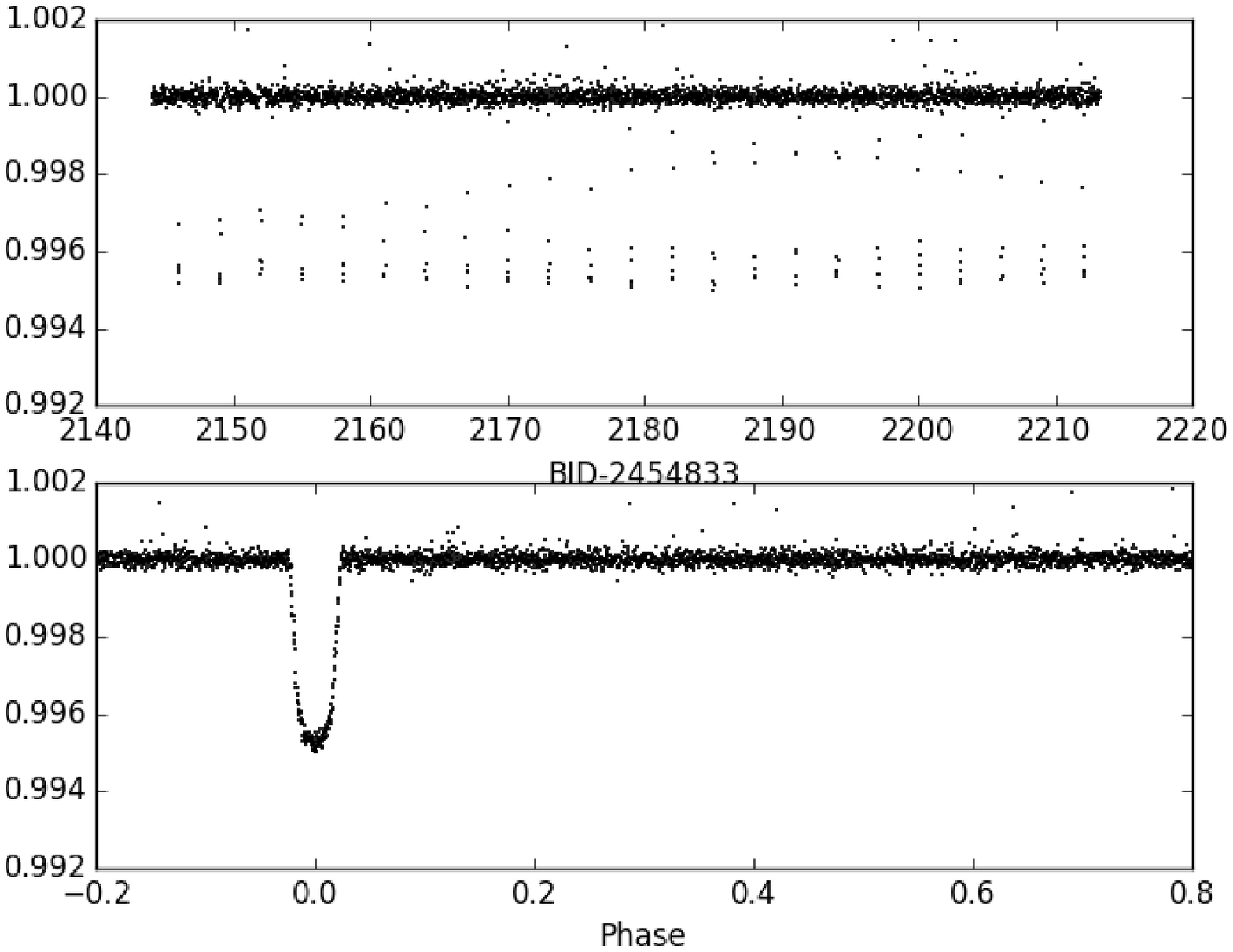}}
\caption{Corrected and normalized light curve of \snameone. The upper plot shows the normalized light curve over time. The lower plot displays the phase folded light curve.\label{fig2}}
\resizebox{0.5\textwidth}{!}{\plotone{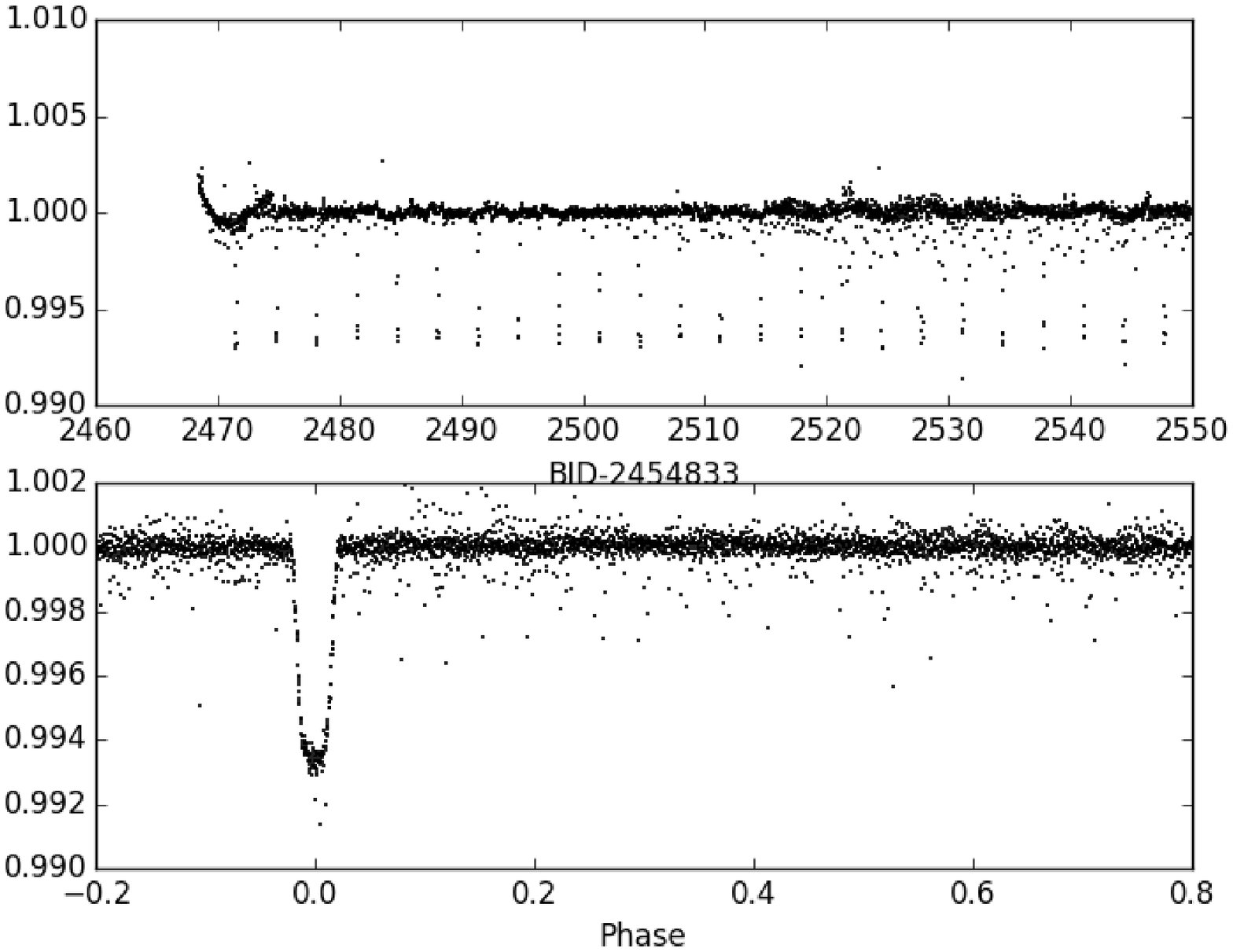}}
\caption{Corrected and normalized light curve of \snametwo. Notation as in Figure \ref{fig2}.\label{fig3}}
\end{figure}

To detect transit signals in K2 campaign 3 and 7, we used the light curves extracted by \citet{Vanderburg2014} from the K2 data. We used the same algorithms and vetting tools described in \citet{cabrera2012} and \citet{Grziwa12,Grziwa16b}. These algorithms have been largely used by our team to detect and confirm planets in other K2 fields \citep{Barragan2016,Grziwa16,Johnson2016,Smith2016}. For the modeling of the transit light curves we used our own optimized photometry employing a similar approach as in \citet{Vanderburg2014}, which allowed us to reduce strong systematics by choosing optimal segment sizes when splitting the light curve for de-correlation. The photometry was performed using a fixed aperture for each object as shown in Fig.~\ref{fig1}. For \snameone\ we selected an aperture of 33 pixels as the star is isolated. In the case of \snametwo, this target is in a field that is close to the galactic center and thus very crowded. We minimized the contamination effects arising from nearby sources by using an fixed aperture of only 9 pixels (Fig.~\ref{fig1}). As in the pipeline of Kepler and \citet{Vanderburg2014}, each light curve was split in segments to remove correlated noise. The length of these segments influences the quality of de-correlation. We found an optimal size for the segments to be twice the orbital period of the planet. This way we avoided splitting the light curve within any transit signal. These short segments were individually de-correlated against the relative motion of the star, given in the \texttt{POS\_CORR} columns\footnote{Due to strong correlation between \texttt{POS\_CORR1} and \texttt{POS\_CORR2}, it was sufficient to use \texttt{POS\_CORR1} for de-correlation.}. To remove long term trends we de-correlated these segments also in the time domain after ruling out the existence of ellipsoidal variations in the phase folded light curve that might hint at eclipsing binary systems. The resulting light curves, in the time domain and phase folded, are shown in Figures~\ref{fig2} and \ref{fig3}.

\subsection{High Dispersion Spectroscopy}
\label{RV_Follow_up}

We acquired five and eight high-resolution spectra ($R$\,$\approx$\,67,000) of \snameone\ and \snametwo\ with the the FIbre-fed \'Echelle Spectrograph \citep[FIES;][]{Frandsen1999,Telting2014} between June and September 2016. FIES is mounted at the 2.56m Nordic Optical Telescope (NOT) of Roque de los Muchachos Observatory (La Palma, Spain). We adopted the same observing strategy as in  \citet{Buchhave2010} and \citet{Gandolfi2015}, i.e., we bracketed each science observation with long exposed ThAr spectra ($T_\mathrm{exp}$\,$\approx$\,35\,sec). The exposure time was set to 1800\,--\,3600~sec -- according to sky conditions and scheduling constraints -- leading to a signal-to-noise ratio (S/N) of 25--35 per pixel at 5500\,\AA. The FIES data were reduced using standard \texttt{IRAF} and \texttt{IDL} procedures. Radial velocity measurements were extracted via multi-order cross-correlation with the RV standard stars \object{HD\,50692} and \object{HD\,182572} \citep{Udry1999} observed with the same instrument set-up as the target stars.

We also took three additional high-resolution spectra of \snameone\ in July 2016 with the HARPS-N spectrograph \citep[R\,$\approx$\,115,000;][]{Cosentino2012} mounted at the 3.58m Telescopio Nazionale Galileo (TNG) at Roque de los Muchachos Observatory (La Palma, Spain). The exposure times were set to 1200\,--\,1500 seconds leading to a S/N of 15--20 per pixel at 5500\,\AA\ for the extracted spectra. We used the second fiber to monitor the Moon background and reduced the data with the HARPS-N dedicated pipeline. Radial velocities were extracted by cross-correlating the extracted spectra with a G2 numerical mask \citep{Baranne96,Pepe02}.

The FIES and HARPS-N RVs and their uncertainties are listed in Table~\ref{rvs}, along with the bisector span (BIS) of the cross-correlation function (CCF). Time stamps are given in barycentric Julian day in barycentric dynamical time (BJD$_\mathrm{TDB}$). 

We searched for possible correlation between the RV and BIS measurements that might unveil activity-induced RV variations and/or the presence of blended eclipsing binary systems \citep{Queloz01}. The Pearson correlation coefficient between the RV and BIS measurements of \snameone\ is 0.11 with a p-value of 0.79. For \snametwo\ the Pearson correlation coefficient is 0.10 with a p-value of 0.81. Adopting a threshold of 0.05 for the p-value confidence level \citep{Lucy1971}, the lack of significant correlations between the RV and BIS measurements of both stars further confirm that the observed Doppler variations are  induced by the orbiting planets.

\begin{table}
\caption{FIES and HARPS-N RV measurements of \snameone\ and \snametwo.\label{rvs}}
\begin{tabular}{lccrr}
\hline
\hline
BJD$_\mathrm{TDB}$ & RV & $\sigma_{\mathrm{RV}}$ &  BIS & Instr. \\
$-$2,450,000 & (\kms) & (\kms)  & (\kms) & \\
\hline
\noalign{\smallskip}
\multicolumn{2}{l}{\texttt{\snameone}} \\
7568.72048 & $-$45.505 & 0.012 &    0.007 & FIES    \\
7569.72143 & $-$45.394 & 0.024 & $-$0.006 & FIES    \\
7570.71124 & $-$45.490 & 0.012 &    0.022 & FIES    \\
7577.71171 & $-$45.532 & 0.012 &    0.014 & FIES    \\
7578.64730 & $-$45.422 & 0.030 &    0.018 & FIES    \\
7585.67140 & $-$45.324 & 0.010 & $-$0.036 & HARPS-N \\
7586.68055 & $-$45.362 & 0.006 & $-$0.023 & HARPS-N \\
7587.70160 & $-$45.260 & 0.007 &    0.007 & HARPS-N \\
\hline
\noalign{\smallskip}
\multicolumn{2}{l}{\texttt{\snametwo}}\\
7565.58753 &  $-$8.276 & 0.025 &    0.069 & FIES \\
7566.56965 &  $-$8.404 & 0.023 &    0.012 & FIES \\
7567.57489 &  $-$8.438 & 0.016 &    0.025 & FIES \\
7568.59817 &  $-$8.272 & 0.021 &    0.044 & FIES \\
7570.54490 &  $-$8.465 & 0.017 &    0.043 & FIES \\
7628.44921 &  $-$8.273 & 0.029 & $-$0.007 & FIES \\
7637.39240 &  $-$8.392 & 0.016 &    0.004 & FIES \\
7640.40979 &  $-$8.429 & 0.016 &    0.035 & FIES \\
\hline
\end{tabular}
\end{table}

\subsection{Imaging}

Imaging with spatial resolution higher than that of K2 is used to detect potential nearby eclipsing binaries that could mimic planetary transit-like signals. It also enables us to measure the fraction of contaminating light arising from potential unresolved nearby sources whose light leaks into the photometric mask of K2, thus diluting the transit signal. \citet{Schmitt2016} observed \snameone\ using the adaptive optics facility at the KECK telescope. They excluded faint contaminant stars as close as $0\farcs25$ up to 4 magnitudes fainter than the target star.

\begin{figure}
\epsscale{0.9}
%\resizebox{\linewidth}{!}{\plotone{f5}}
\plotone{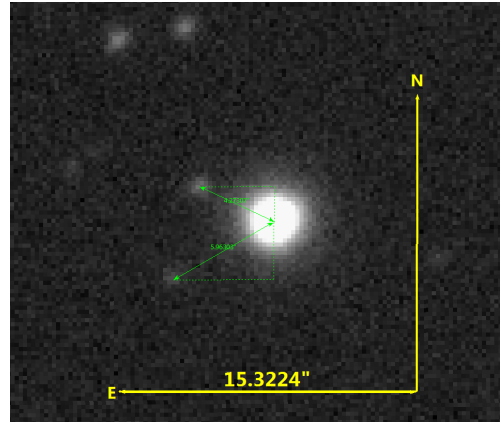}
\caption{ALFOSC@NOT R-band image of \snametwo. We can resolve sources as close as $2\arcsec$ to our target star. \snametwo\ and its two contaminants are marked with green circles.\label{fig5}}
\end{figure}

We observed \snametwo\ on 2016 September 13 (UT) with the ALFOSC camera mounted at the Nordic Optical Telescope. We used the Johnson's standard R-band filter and acquired 16 images of 6 sec and 2 images of 20 sec. The data were bias subtracted and flat-fielded using dusk sky flats. The co-added 6-sec ALFOSC exposures are shown in Figure~\ref{fig5}. We detected two nearby faint stars located $4.3\arcsec$ North-East and $6.0\arcsec$ South-East of \snametwo. They are 6.3 and 6.5 magnitudes fainter than the target and fall inside the photometric aperture that we used to extract the light curve of \snametwo\ from the K2 images. We measured a contribution of $0.005\pm0.001$ to the total flux, by contaminating sources for \snametwo. Our observations exclude additional contaminants out to a separation of $2\arcsec$ and up to 6 magnitudes fainter than the target. We compared our findings with the first data release from GAIA \citep{GAIA_DR1} and found a contamination factor of 0.0043, in agreement with our estimate. No additional sources are present in the GAIA catalog \citep{Lindegren2016} within a radius of $10\arcsec$. The resolving power of GAIA is well below $1\arcsec$.  

\section{Analysis}

\subsection{Spectral Analysis}

We derived the spectroscopic parameters of \snameone\ and \snametwo\ from the co-added spectra used to extract the RVs of the stars (Sect.~\ref{RV_Follow_up}). The stacked FIES and HARPS-N have a S/N of 62 and 32 per pixel at 5500~\AA; the co-added FIES data of \snametwo\ have a S/N of 76 per pixel at 5500~\AA. The analysis was carried out in three independent ways. %In common among the methods is that they use the same observed spectra and fit theoretical models to this, using spectral features that are sensitive to different parameters.

The first technique uses \texttt{ATLAS~9} model spectra \citep{Castelli2004} to fit spectral features that are sensitive to different photospheric parameters. We adopt the calibration equations of \citet{Bruntt2010} and \citet{Doyle2014} to determine the microturbulent (\vmic) and macroturbulent (\vmac) velocities. We mainly used the wings of the H$_\alpha$ and H$_\beta$ lines to estimate the effective temperature ($T_\mathrm{eff}$), and the Mg\,{\sc i}~5167, 5173, and 5184~\AA, Ca\,{\sc i}~6162 and 6439~\AA, and the Na\,{\sc i}~D lines to determine the surface gravity \logg. We simultaneously fit different spectral regions to measure the metal abundance [M/H]. The projected rotational velocity \vsini\ was determined by fitting the profile of many isolated and unblended metal lines.

For the second method, micro-turbulent (\vmic)~and macroturbulent (\vmac)~velocities, as well as the projected stellar rotational velocity \vsini\ were determined as described above. For the spectral analysis the second  method relies on the use of the package SME (Spectroscopy Made Easy, where we used version 4.43) \citep{Valenti1996,Valenti2005}. SME  calculates, using a grid of models (we used the Atlas 12) for a set of given stellar parameters, synthetic spectra of stars and fits them to the observed high-resolution spectra using a $\chi$-square minimizing procedure.

The third method uses the equivalent width (EW) method to derive stellar atmospheric parameters: {\it i}\,) \teff\ is measured by eliminating trends between abundance of the chemical elements and the respective excitation potentials; {\it ii}\,) \logg\ is derived by assuming the ionization equilibrium condition, i.e. requiring that for a given species, the same abundance (within the uncertainties) is obtained from lines of two ionization states (typically, neutral and singly ionized lines); {\it iii}\,) microturbulent velocity is set by minimizing the slope of the relationship between abundance and the logarithm of the reduced EWs. We measured the equivalent widths using the \texttt{DOOp} program {Cantat2014}, a wrapper of \texttt{DAOSPEC} \citep{Stetson2008}. We derived the photospheric parameters with the program \texttt{FAMA} \citep{Magrini2013}, a wrapper of \texttt{MOOG} \citep{Sneden2012}. The adopted atomic parameters are the public version of those prepared for the Gaia-ESO Survey \citep{Heiter2015} and based on the VALD3 data \citep{Ryabchikova2011}. We typically used 200 Fe\,{\sc i} lines and 10 Fe\,{\sc ii} lines for the determination of stellar parameters.

The three methods provide consistent results within two sigma (see Table \ref{tab:stellar1}). The final adopted values are the weighted mean of the three independent determinations, using the error bars to calculate the weighting factor. The stellar parameters for both systems are listed in Table~\ref{tab:stellar}, along with the main identifiers and optical and near-infrared magnitudes.

\begin{table*}[!th]
\begin{center}
\caption{Effective temperature, surface gravity, and metallicity from different spectral analysis methods.\label{tab:stellar1}}
\begin{tabular}{l|ccc|ccc}
\hline
\hline
\noalign{\smallskip}
&& \snameone && & \snametwo &\\
Method &  \teff\ (K) & \logg\ (cgs) &  [Fe/H] (dex) &  \teff\ (K) & \logg\ (cgs) &  [Fe/H] (dex) \\
\noalign{\smallskip}
\hline
\noalign{\smallskip}
Method 1 & $5480 \pm 85 $ & $4.05 \pm 0.15$ & $-0.10 \pm 0.10$ & $6050 \pm 110$ & $3.95 \pm 0.10$ & $0.08 \pm 0.06$\\
Method 2 & $5350 \pm 90 $ & $3.95 \pm 0.10$ & $-0.10 \pm 0.08$ & $5970 \pm 100$ & $4.30 \pm 0.15$ & $0.08 \pm 0.08$\\
Method 3 & $5625 \pm 115$ & $4.22 \pm 0.07$ & ~~$0.24  \pm 0.15$ & $6080 \pm 150$ & $3.95 \pm 0.05$ & $0.13 \pm 0.16$\\
\noalign{\smallskip}
\hline
\end{tabular}
\end{center}
\end{table*}

\begin{table*}[!th]
\begin{center}
\caption{Main identifiers, coordinates, magnitudes, and spectroscopic parameters of both systems.\label{tab:stellar}}
\begin{tabular}{lccc}
\hline
\hline
\noalign{\smallskip}
Parameter & {\texttt{\snameone}} &  {\texttt{\snametwo}} & Unit\\
\noalign{\smallskip}
\hline
\noalign{\smallskip}
RA 	& 22$^h$34$^m$25$^s$.49  & 18$^h$59$^m$56$^s$.49  & h \\
DEC & -13$\degr$43$^\prime$54$^{\prime \prime}$.13 & -22$\degr$17$^\prime$36$^{\prime \prime}$.25 & deg\\
2MASS ID & 22342548-1343541 & 18595649-2217363 & \ldots\\
EPIC ID & 206038483 & 216468514 & \ldots\\
\noalign{\smallskip}
\hline
\noalign{\smallskip}
Effective Temperature \teff & \steffone & \stefftwo & K\\
Surface Gravity \logg & \sloggone & \sloggtwo & cgs\\
Metallicity [Fe/H] & \smetone   & \smettwo & dex \\
%\vmic & $0.95 \pm 0.03$   & \kms &\\
\vsini & $2.2 \pm 0.5$& $4.6 \pm 0.5$& \kms \\
Spectral Type & G4\,V & F9\,IV & \ldots\\
\noalign{\smallskip}
\hline
\noalign{\smallskip}
B mag (UCAC4) & $13.56 \pm 0.01$ & $13.64 \pm 0.01$ & mag\\
V mag (UCAC4) & $12.79 \pm 0.02$ & $12.92 \pm 0.01$ & mag\\
J mag (2MASS)& $11.41  \pm 0.02$ & $11.56 \pm 0.02$ & mag\\
H mag (2MASS)& $11.09  \pm 0.03$ & $11.26 \pm 0.03$ & mag\\
K mag (2MASS)& $10.99  \pm 0.02$ & $11.21 \pm 0.02$ & mag\\
\noalign{\smallskip}
\hline
\end{tabular}
\end{center}
\end{table*}

\subsection{Joint Analysis of Photometric and Radial Velocity Measurements}

We used the Transit Light Curve Modelling (\texttt{TLCM}) code  (\citealt{Csizmadia2015}; Csizmadia  et  al.  in  prep.) for the simultaneous analysis of the detrended light curves and radial velocity measurements. \texttt{TLCM} uses the \cite{M&A} model to fit planetary transit light curves. The RV measurements are modeled with a Keplerian orbit. The fit is optimized using first a genetic algorithm and then a simulated annealing chain.

The fitted parameters are the semi-major axis $a/R_*$ and planet radius $R_\mathrm{p}/R_*$, both scaled to the radius of the star, the orbital inclination $i$, the limb darkening coefficients $u_+=u_1+u_2$ and $u_- = u_1 -u_2$, the radial velocity semi amplitude $K$ and the systemic $\gamma$-velocity. The period ($P_\mathrm{orb}$) and epoch of mid-transit ($T_0$) are allowed to vary slightly around the values determined already by the detection.

For \pnametwo\ the model did not converge to the global minimum when leaving all nine parameters completely free, instead it seemed to converge to a broader local minimum. We thus first modeled the light curve, keeping the epoch and period, as well as the limb darkening coefficients fixed using estimates from \citet{Claret2011}. This gave us first estimates on the inclination, planet radius ratio, and semi major axis. In a second step we fitted all nine free parameters as for \pnameone, but restricted the parameter space with the priors as given by our first fit. To verify our results, We also modeled the light curve with different fixed inclinations, leaving all other parameters free. This confirmed our result of an high impact parameter.

We also fit the data for non-circular orbits. The best fitting eccentricity for \snameone\ is  $0.09 \pm 0.03$ with a p-value of 0.90; as for \snametwo, we obtained $0.06 \pm 0.05$ with a p-value of 0.57. Both p-values are larger than the 0.05 level of significance suggested by \citet{Lucy1971}. We concluded that the RV measurements do not allow us to prefer the eccentric solutions over the circular ones and thus fixed the orbit eccentricities to zero. This assumption is reasonable given the fact that short period orbits are expected to have circularized. Using the equations from \citet{Leconte2010}, we calculated the tidal time-scales for the eccentricity evolution of the two systems\footnote{The rotation periods of the stars are estimated from the stellar radii and \vsini, assuming that the objects are seen equator-on.}. Assuming a modified tidal quality factor of $Q^\prime_\star=10^{6.5}$ for the stars and $Q^\prime_\mathrm{p}=10^{5.5}$ for the planets \citep{Jackson2008}, the timescales are $\sim$400 and $\sim$25~Myr for \snameone\ and \snametwo, respectively. These time scales are shorter than the estimated ages of the two host stars (Table~\ref{tab:param}).

We also fitted for radial velocity trends that might unveil the presence of additional orbiting companions in the systems. We obtained radial accelerations that are consistent with zero.

\begin{figure}
\resizebox{.5\textwidth}{!}{\plotone{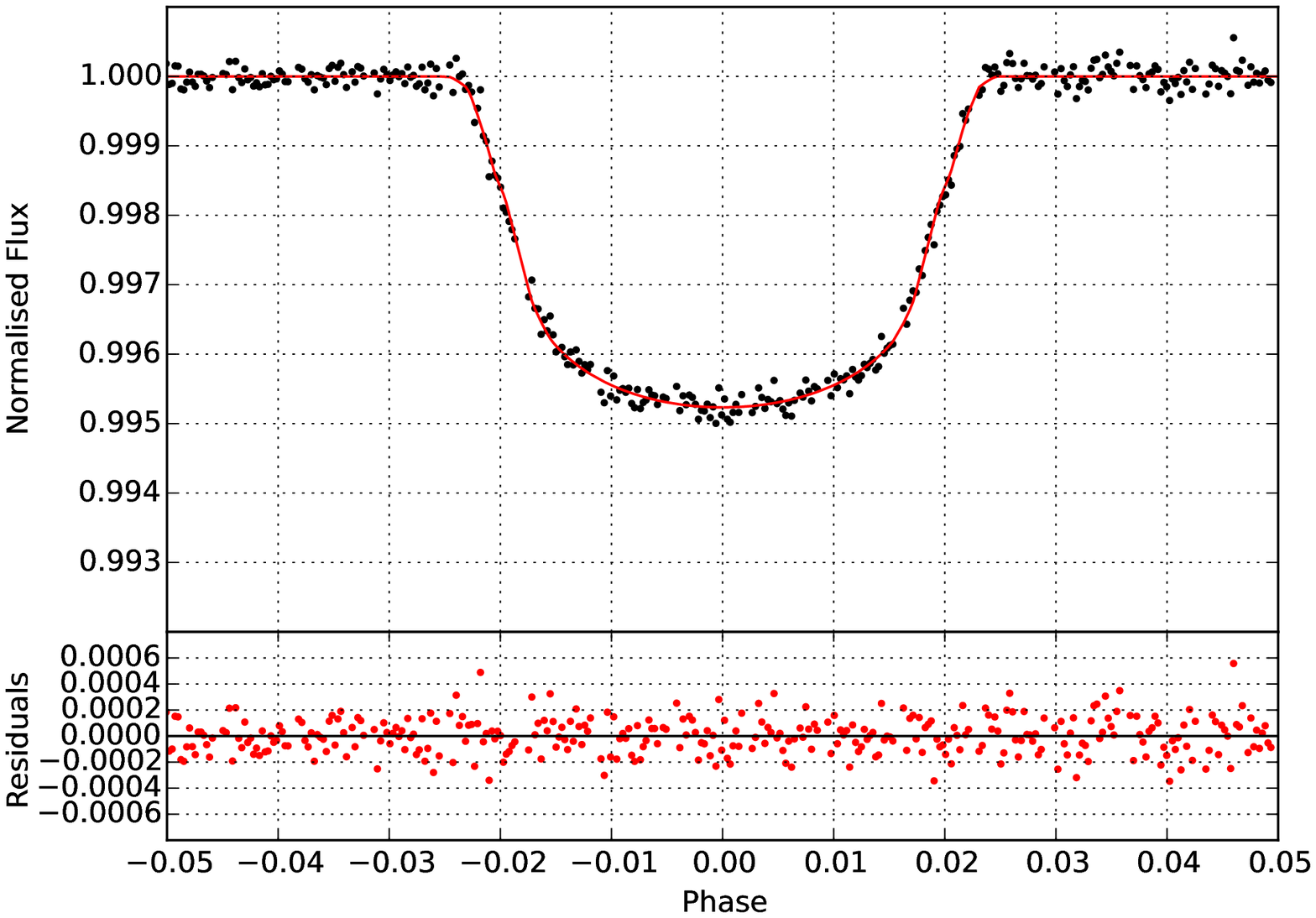}}
\caption{Phase folded light curve and best fitting transit model (red line) of \pnameone. Residuals to the fit are shown in the lower panel.\label{fig6}}
\resizebox{.5\textwidth}{!}{\plotone{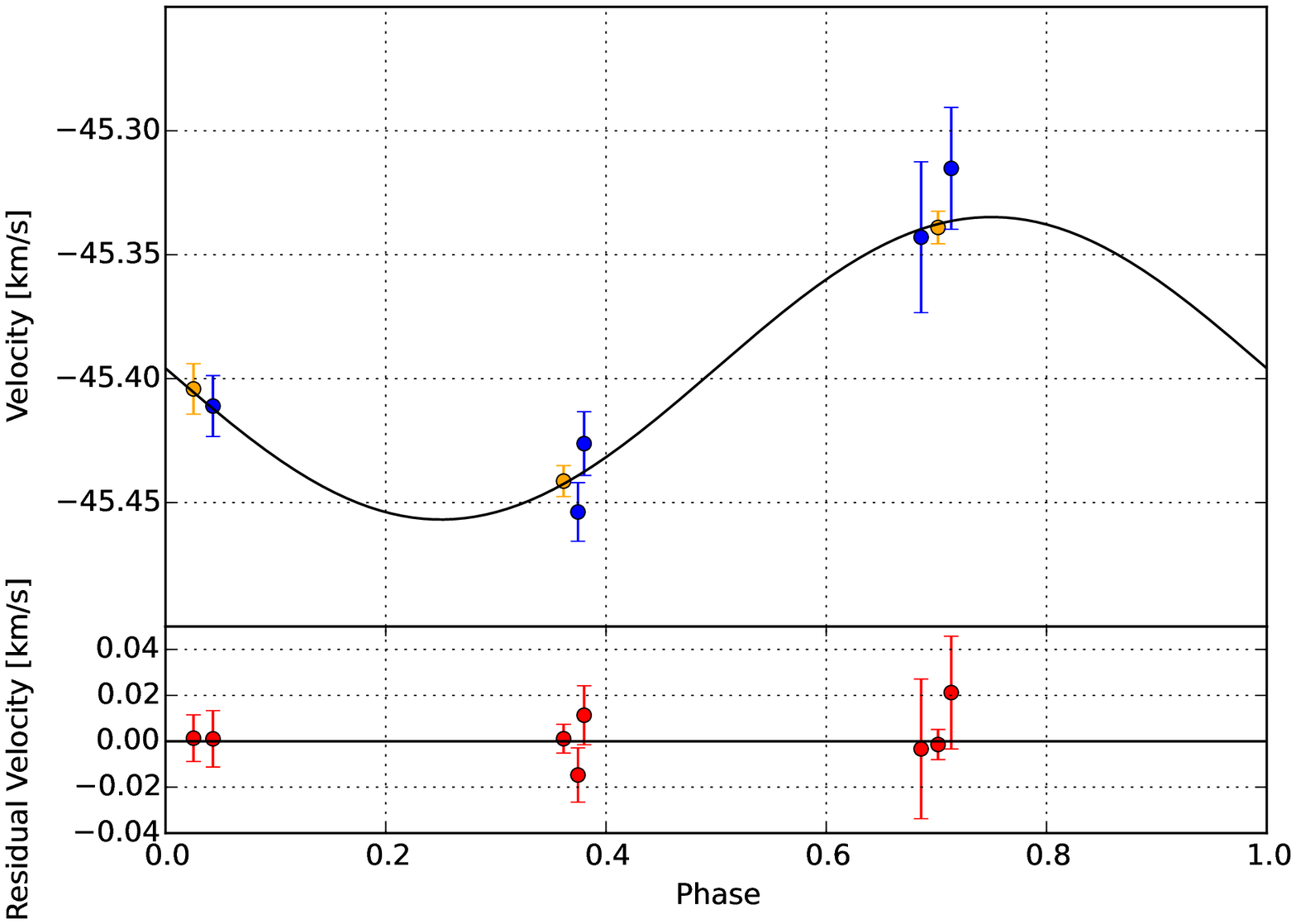}}
\caption{FIES (blue circles) and HARPS-N (orange circles) RV measurements of \pnameone\ and best fitting circular model. Residuals to the fit are shown in the lower panel.\label{fig7}}
\end{figure}

\begin{figure}
\resizebox{.5\textwidth}{!}{\plotone{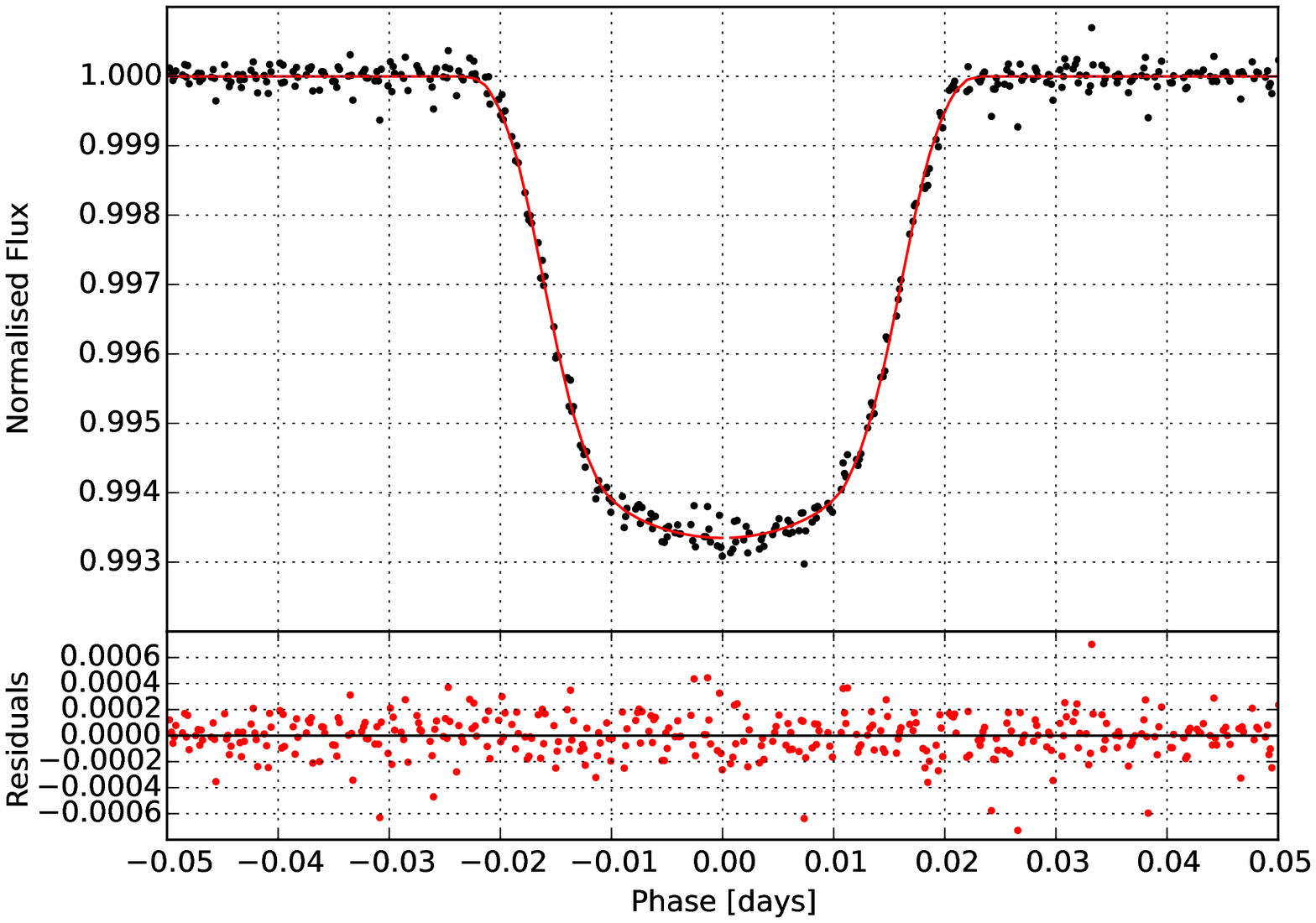}}
\caption{Phase folded light curve and the best fitting transit model (red line) of \pnametwo. Residuals to the fit are shown in the lower panel.\label{fig8}}
\resizebox{.5\textwidth}{!}{\plotone{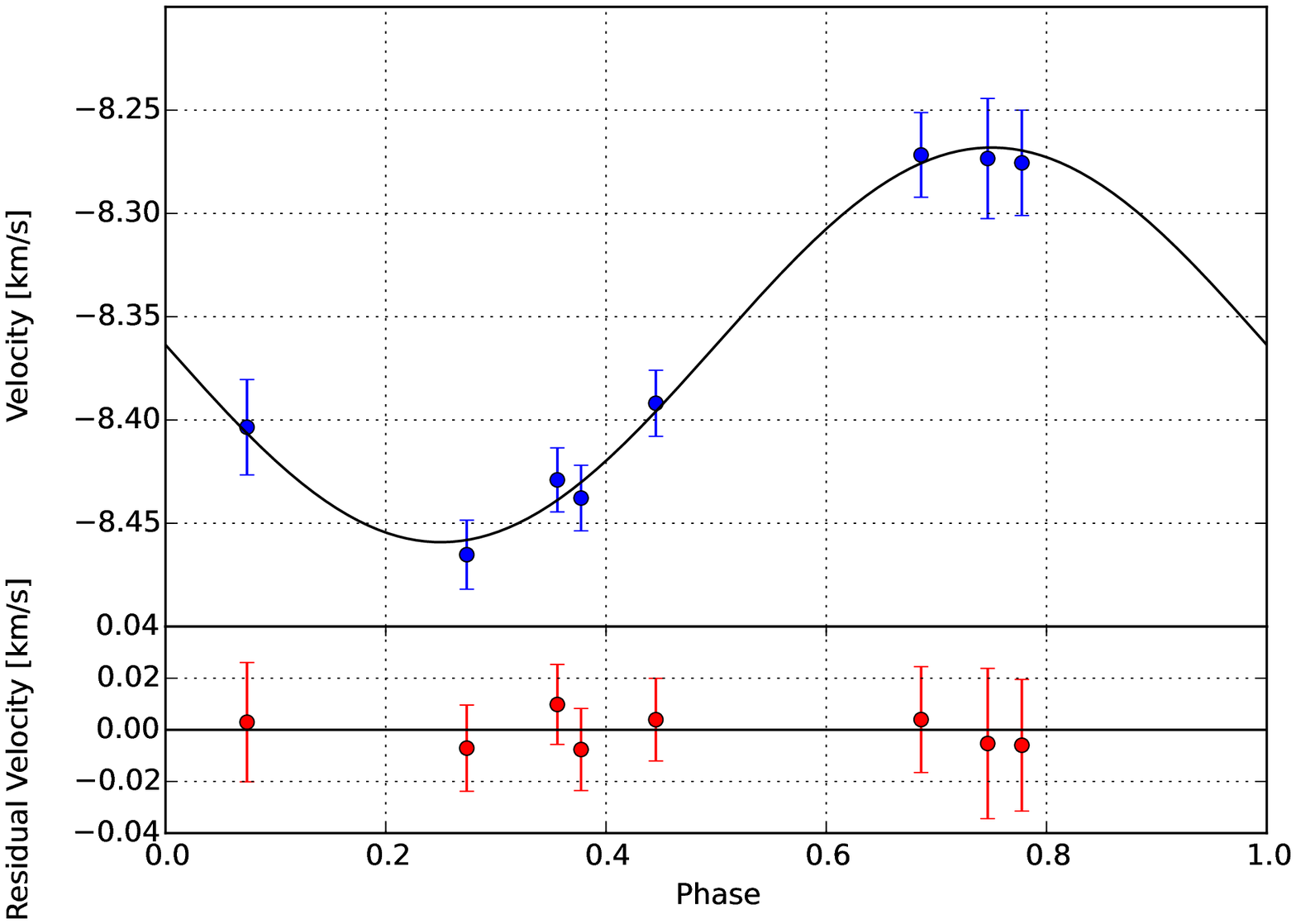}}
\caption{FIES RV measurements of \pnametwo\ and best fitting circular model. Residuals to the fit are shown in the lower panel.\label{fig9}}
\end{figure}

The best fitting transit model and circular RV curve of \pnameone\ are shown in Figures~\ref{fig6} and \ref{fig7}, along with the photometric and RV data. Results for \pnametwo\ are displayed in Figures~\ref{fig8} and \ref{fig9}. We checked our results by performing a joint fit to the photometric and RV data using the MCMC code \texttt{pyaneti} \citep{Barraganprep}. Following the same method outlined in \citet{Barragan2016}, we set uninformative uniform priors in a wide range for each parameter and explored the parameter space with 500 chains. The final parameter estimates are consistent within 1-$\sigma$ with those obtained using \texttt{TLCM}.

From the results of the spectral analysis and joint data modeling, we used Yonsei-Yale \citep{Yi2001,Demarque2004} and Dartmouth  \citep{Dotter2008} isochrones to estimate masses, radii, and ages of \snameone\ and \snametwo. We obtained results that are in agreement regardless of the adopted set of isochrones. For the final results we used the Yonsei-Yale isochrones \citep{Yi2001,Demarque2004}. From the fundamental parameters of the host stars we calculated radii and masses of the two transiting planets. The parameter estimates are listed in Table~\ref{tab:param} for both systems.

\section{Discussion and Summary}
\subsection{\pnameone}
\pnameone\ is a transiting sub-Jovian planet with an orbital period of $3.00267 \pm 0.00006$ ~days. It orbits a G4 main sequence star. The planet's calculated effective temperature is \pteffone ~K. With a radius of \pradone~\Rjup\ and a mass of \pmassone~\Mjup, it is more dense than expected. The radius anomaly, based on the difference between model estimated to observed radius as described in \citet{Laughlin2011}, is $-0.46$, making this planet more dense than expected.  Adaptive optics imaging by \citet{Schmitt2016} shows that there is no light contamination that could cause an underestimation of the planetary radius. We can exclude ellipsoidal variation with amplitudes above 0.05\,mmag in the light curve. There is no obvious trend in the radial velocity data, although we can not exclude radial accelerations lower than 0.002~km~s$^{-1}$day$^{-1}$.

\begin{figure*}
\plotone{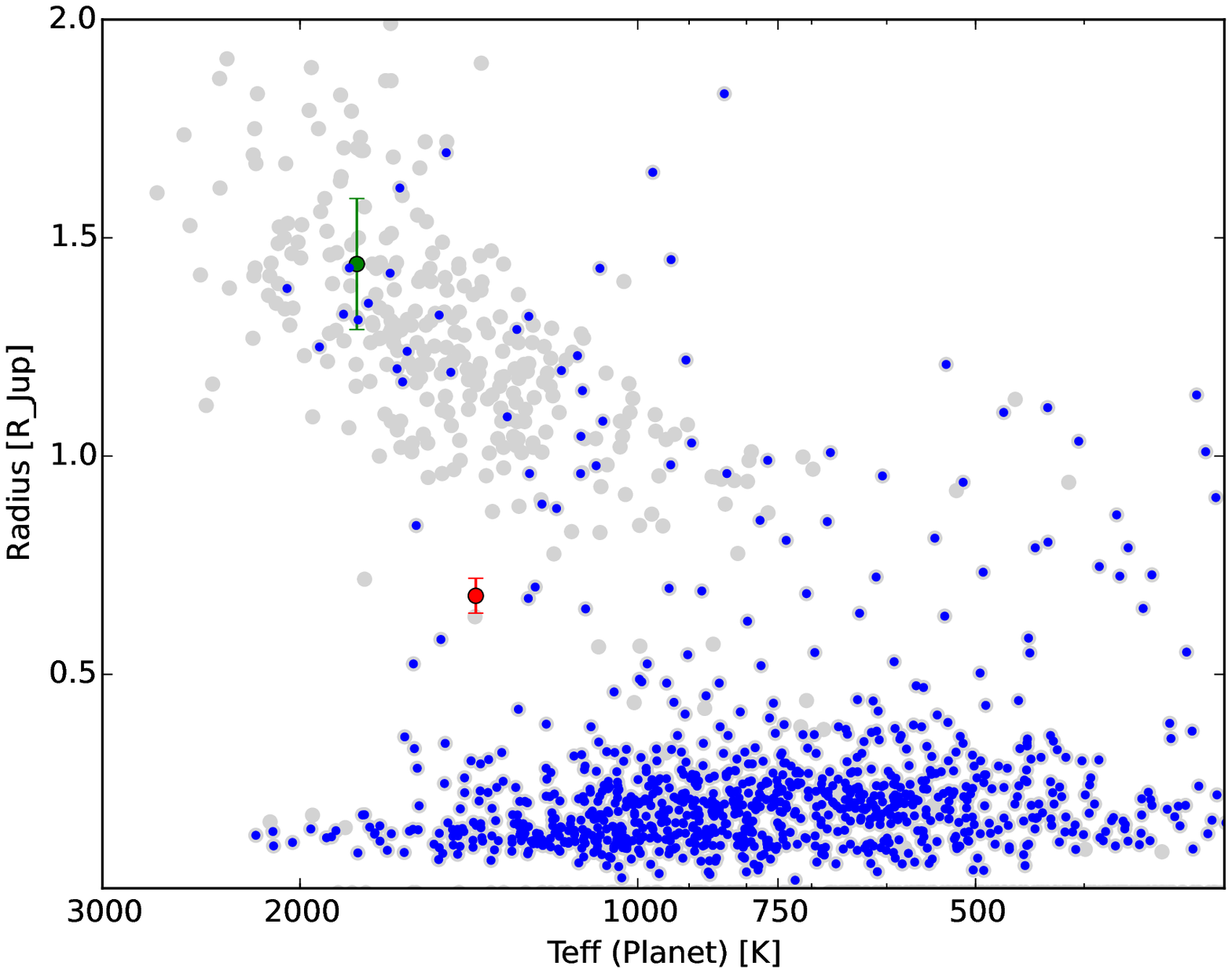}
\caption{Planet radius over its orbit-averaged effective temperature. The gray dots show all planets. The blue dots mark planets that have been detected by the Kepler spacecraft (Kepler mission or K2 mission). The red dot denotes \pnameone\ and the green dot \pnametwo. The exoplanet data are taken from Extrasolar Planets Encyclopaedia (\url{www.exoplanets.eu}).\label{fig11}}
\end{figure*}

The short orbital period and high effective temperature of the planet, along with its sub-Jovian size, put \pnameone\ close to the the so called sub-Jovian desert. Fig.~\ref{fig11} shows the known transiting planets with their radii plotted against their calculated effective temperatures as given by the equation in \citet{Laughlin2011}
\begin{equation}
T_{eff} = \left(\frac{R_{S}}{2a}\right)^{1/2} \frac{T_S}{(1-e^2)^{1/8}}.
\end{equation}

There is a clear lack of hot sub-Jovian planets. Due to different observational biases of exoplanet surveys (e.g., most of the inflated hot Jupiters have been detected by ground based surveys, which might not be able to detect sub-Jovian planets with the same efficiency) the upper border is not as well defined as it may seem. This can be seen by looking only at confirmed planets of the Kepler spacecraft (blue points in Fig.~\ref{fig11}). Nevertheless, all observations suggest that the sub-Jovian desert exists, although its borders are not well defined. Only a few planets are known in this regime \citep[e.g.][]{Sato2005, Bonomo2014}. \pnameone\ might help in the future to get better restrictions on its borders.

\subsection{\pnametwo}
\pnametwo\ is a Jovian planet on a short orbital period of $3.31392\pm 0.00002$ days. The planet orbits a F9 star about to leave the main sequence. It is one of only a few transiting planets known to orbit sub giants \citep[e.g.][]{Smith2016, Pepper2016, Eylen2016, Almenara2015}. The planet's calculated effective temperature is \ptefftwo ~K its radius is \pradtwo~\Rjup\ and its mass is \pmasstwo~\Mjup. The radius anomaly is $+0.21$, making \pnametwo\ in contrast to \pnameone\ an highly inflated gaseous planet. Such high inflation has already been observed for other giant planets with a similar effective temperature (see Figure \ref{fig11}). As suggested by \citet{Laughlin2011}, Ohmic heating might be at least partly responsible for such inflation of the planet.

Since it is projected against the galactic center, \snametwo\ is in a relatively crowded stellar region. Using seeing-limited imaging and the GAIA public archive (DR1) we identify two faint stars within $\sim$$10\arcsec$. The resulting contamination factor of 0.005 has been taken into account when modeling the light curve. The radial velocity data do not show any significant eccentricity or long term trend higher than 0.001~km s$^{-1}$day$^{-1}$.
The light curve of \snametwo\ shows no ellipsoidal variation with an amplitude larger than 0.1~mmag.

\begin{table*}[!th]
\begin{center}
\caption{Parameters from light curve and RV data analysis.\label{tab:param}}
\begin{tabular}{lccc}
\hline
\hline
\noalign{\smallskip}
Parameter & \texttt{\snameone} & {\texttt{\snametwo}}& Unit\\
\noalign{\smallskip}
\hline
\noalign{\smallskip}
Orbital period $P_\mathrm{orb}$  & $3.00265 \pm 0.00004$		& $3.31392 \pm  0.00002$	& days\\
Transit epoch $T_0$ 			 & $6928.0593 \pm 0.0007$ 		& $7304.5244 \pm 0.0002$ 		& BJD$_\mathrm{TDB}-2450000$\\
Transit Duration 				 & $3.08 \pm 0.10$				& $3.19 \pm 0.45$ & hours\\
Scaled semi-major axis $a/R_*$ 	 & $7.78 \pm 0.17$   		& $5.75 \pm 0.31$ & \\
Semi-major axis $a$ 			 & $ 0.045 \pm 0.003$   	& $0.048 \pm 0.005$ & au\\
Scaled planet radius $R_P/R_*$ 	 & $0.063 \pm 0.001$  & $0.083 \pm 0.001$& \\
Orbital inclination angle $i$	 & $88.49 \pm 0.96$       & $81.9 \pm 0.7$& $\deg$\\
Impact parameter $b$ 			 & $0.21 \pm 0.13$                & $0.81 \pm 0.08$& \\
Limb-darkening coefficient $u_+$ & $0.63 \pm 0.08$    & $0.51 \pm 0.08$& \\
Limb-darkening coefficient $u_-$ &  $0.29 \pm 0.10$   & $0.17 \pm 0.07$& \\
\noalign{\smallskip}
\hline
\noalign{\smallskip}
Radial velocity semi amplitude $K$ &  $61.0 \pm 2.6$    & $95.5 \pm 1.3$     & $ \ms $\\
Systemic radial velocity $\gamma$ & $-45.475 \pm 0.003$ & $-8.364 \pm 0.001$ & \kms  \\
RV velocity offset between FIES and HARPS &      $0.158 \pm 0.004$       & - & \kms \\
Eccentricity $e$& 0 (fixed)                         & 0 (fixed) & \\
\noalign{\smallskip}
\hline
\noalign{\smallskip}
Stellar mass $M_*$     			& \smassone          & \smasstwo & \Msun\\
Stellar radius $R_*$   			& \sradone           & \sradtwo & \Rsun\\
$M^{1/3}_*/R_*$  					& $0.89 \pm 0.02$                   & $0.61\pm 0.03$& Solar units\\
Stellar mean density $\rho_*$       			& $0.99 \pm 0.06$ & $0.33 \pm 0.05$ & \gcm \\
Stellar surface gravity \logg\footnote{\label{note}Derived from the light curve modeling, effective temperature, metal content, and isochrones.} & \slogglcone  & \slogglctwo & cgs\\
Age                    			& \sageone               & \sagetwo & Gyr\\
%Distance               & & & parsec\\
\noalign{\smallskip}
\hline
\noalign{\smallskip}
Planetary mass $M_P$   			& \pmassone & \pmasstwo & $M_\mathrm{Jup}$\\
Planetary radius $R_P$ 			& \pradone & \pradtwo & $R_\mathrm{Jup}$ \\
Planetary mean density          & $1.7 \pm 0.3$ & $0.35 \pm 0.1$ & cgs\\
Planetary surface gravity \loggp         			& \ploggone & \ploggtwo & cgs\\
Planetary calculated effective temperature & \pteffone & \ptefftwo& K\\
\noalign{\smallskip}
\hline
\end{tabular}
\end{center}
\end{table*}

\acknowledgments

We would like to express our deepest gratitude to the NOT and TNG staff members for their unique support during the observations and scheduling of our runs. Szilard Csizmadia thanks the Hungarian OTKA Grant K113117. Hans Deeg and David Nespral acknowledge support by grant ESP2015-65712-C5-4-R of the Spanish Secretary of State for R\& D\&i (MINECO). This research was supported by the Ministerio de Economia y Competitividad under project FIS2012-31079. The research leading to these results has received funding from the European Union Seventh Framework Programme (FP7/2013-2016) under grant agreement No.~312430 (OPTICON) and from the NASA K2 Guest Observer Cycle 1 program under grant NNX15AV58G to The University of Texas at Austin. Based on observations obtained with the Nordic Optical Telescope (NOT), operated on the island of La Palma jointly by Denmark, Finland, Iceland, Norway, and Sweden, in the Spanish Observatorio del Roque de los Muchachos (ORM) on the island of La Palma, and with the Italian Telescopio Nazionale Galileo (TNG) operated also at the ORM (IAC) by the INAF - Fundaci\'on Galileo Galilei. The data presented here were obtained in part with ALFOSC, which is provided by the Instituto de Astrofisica de Andalucia (IAA) under a joint agreement with the University of Copenhagen and NOTSA. This work has made use of data from the European Space Agency (ESA) mission {\it Gaia} (\url{http://www.cosmos.esa.int/gaia}), processed by the {\it Gaia} Data Processing and Analysis Consortium (DPAC, \url{http://www.cosmos.esa.int/web/gaia/dpac/consortium}). Funding for the DPAC has been provided by national institutions, in particular the institutions participating in the {\it Gaia} Multilateral Agreement.
We are happy to acknowledge the continued involvement with help and upgrades to the Spectroscopy Made Easy (SME) program package, by N Piskunov and J. Valenti. SME makes use of the VALD database, operated at Uppsala University, the Institute of Astronomy RAS in Moscow, and the University of Vienna \citep{Ryabchikova2015}.

\facility{NOT (FIES, ALFOSC), TNG (HARPS-N), Kepler (K2), GAIA}.

\software{\texttt{IDL}, \texttt{IRAF}, \texttt{SME}, \texttt{DOOp}, \texttt{FAMA}, \texttt{TLCM}, \texttt{pyaneti}.}

\end{document}